\documentclass[%
 reprint,
 amsmath,amssymb,
 aps,
 prd,
 superscriptaddress
 longbibilography
]{revtex4-2}

\usepackage{graphicx}
\usepackage{dcolumn}
\usepackage{bm}
\usepackage{threeparttable}
\usepackage{subcaption}
\usepackage{booktabs}
\usepackage[colorlinks=true,
            linkcolor=blue,
            citecolor=blue,
            urlcolor=blue]{hyperref}

\begin{document}

\preprint{APS/123-QED}

\title{Spectroscopic and Structural Properties of $B$, $B_s$, and $B_c$ Mesons within a Non-Relativistic Potential Model: A Comparative Analysis via Matrix Numerov and Variational Methods}
\thanks{Potential model formalism (Killingbeck Potential)}%

\author{Pritom Kumar Nath}
 \ \altaffiliation[Also at ]{Department of Physics, Tezpur University.}
 \email{php24117@tezu.ac.in}
\author{Rishan Dev Choudhury}%
 \email{php24102@tezu.ac.in}
\author{Jugal Lahkar}
 \email{shriyan@tezu.ernet.in}
\affiliation{%
 Tezpur University, Napaam, Tezpur (Assam) \\
 Pin: 784028
}%

\date{\today}

\begin{abstract}
In this work, we studied the spectroscopic masses, decay constants, oscillation frequencies, and Isgur-Wise function parameters of $B$, $B_s$, and $B_c$ mesons within the framework of a non-relativistic potential model. The interaction is modelled using the Killingbeck potential, with calculations performed using two complementary approaches: the matrix Numerov method for solving the Schrödinger equation and the quantum mechanical variational method employing the Gaussian wave function. We determined the unknown parameters of the Killingbeck potential by fitting it to the Cornell potential using the ordinary least squares (OLS) method. These calibrated parameters are then used to compute the aforementioned physical observables.We obtained the masses of the heavy mesons with deviations of less than 0.5\% from the highly precise experimental data. We also discussed the drawbacks and limitations of the non-relativistic model and highlighted its domain of applicability. At the end, we compare our results obtained from both methods with those from other theoretical frameworks, including QCD sum rules, lattice QCD, and alternative potential model studies.
\end{abstract}

\maketitle


\section{\label{sec:level1}Introduction}

The spectroscopy of heavy-flavoured mesons is the foundation of non-perturbative QCD for analysing the strong force in the Standard Model of particle physics. Within this non-perturbative sector\cite{Peskin1995QFT}, mesons containing at least one bottom quark, such as \(B(\bar{b}d)\), \(B_s (\bar{b}s)\), and \(B_c(\bar{b}c)\), are important heavy-light and heavy-heavy mesonic systems. Due to the heavy quark mass, the bottom quark creates a clean separation of physical scales ($\Lambda_{QCD}\approx0.2GeV$)\cite{Neubert2005EFTReview}, which allows us to use these effective field theories to simplify complex interactions. In this work, we adopt the Killingbeck potential\cite{Killingbeck1978}. We solve the radial Schr\"odinger equation numerically using the Matrix Numerov method~\cite{Barnea2026HighOrderMatrixNumerov,Pillai2012MatrixNumerov}, which is well known for its precision of fourth-order global truncation, $\mathcal{O}(\Delta r^4)$, where $\Delta r$ represents the spatial spacing of the grid. The method involves transforming the Schr\"odinger equation into a reduced matrix form, followed by the solution of the corresponding matrix eigenvalue problem in which the kinetic and potential energy operators are discretised on a finite spatial grid. To assess the reliability of the numerical results, a detailed comparison is performed with the variational approach employing a Gaussian trial wave function. Using both methods, we investigate the spectroscopic properties of heavy mesons, including the meson mass, decay constant, and oscillation frequency, together with the structural properties characterized by the parameters of the Isgur--Wise function, namely the slope, $\rho^2$ and curvature, $C$. Furthermore, the comparative analysis provides insight into the sensitivity of these observables to variations in the potential parameters, thereby offering a quantitative understanding of how mesonic properties depend on the underlying interaction potential. 

The Schrödinger equation with the Killingbeck potential is not exactly solvable. Throughout the years, various approximation schemes as well as numerical methods have been used to solve such problems. The most common approaches include the Runge-Kutta method, discrete variable representation (DVR), the matrix Numerov method, and the variational method. In the present work, we employ the matrix Numerov method\cite{Pillai2012MatrixNumerov} for its simplicity and computational efficiency, complemented by the variational method. The latter relies solely on the selection of the trial wavefunction. In our case, we employ a Gaussian basis, which accurately models the localised spatial extent of the bound state while ensuring computational efficiency when evaluating high-order derivative operators\cite{Choudhuryetal2026}.  

Some notable phenomenological potentials include the power-law \cite{M_2023}, Richardson\cite {M_2023}, Cornell, Killingbeck\cite {M_2023}, and logarithmic potentials\cite {M_2023}, which have been employed to study quark-antiquark bound states. In our present work, the unknown parameters of the Killingbeck potential were determined by fitting it to the Cornell potential\cite{Peskin1995QFT}\cite{PhysRevD.21.203}\cite{M_2023} using the well-established Cornell parameters\cite{Kingkar2022Cornell}, which are the strong coupling constant $\alpha_s$, the string tension $b$, and the constant term $C$. We have carried out the fitting employing the OLS method\cite{Press1989NumericalRecipes} to obtain the optimal values of the Killingbeck parameters. It is to be noted that the Cornell potential is widely used in meson spectroscopy. So, the pertinent question is the emergence of the harmonic term $ar^2$.  However, for a heavy-quarkonium bound state interacting with a complex, non-perturbative vacuum, short-distance Debye screening of the vector Coulombic and scalar string interactions yields a Taylor-expanded Karsch-Mehr-Satz (KMS) potential\cite{Karsch1988}. This potential maps directly onto the Killingbeck form given in Eq.(~\ref{eq:killingbeck_general}) (see the Appendix). The choice of the Gaussian trial wave function in the variational method is motivated by the harmonic-oscillator term in the Killingbeck potential. The Cornell potential was chosen as the reference model because of its remarkable success in describing the quark-antiquark interaction. It was originally formulated to reproduce the charmonium mass spectrum and has since become one of the most extensively adopted phenomenological potentials in heavy-meson spectroscopy due to its ability to simultaneously incorporate the short-range Coulombic interaction and the long-range linear confinement. The term $-\frac{4}{3}\frac{\alpha_s}{r}$\cite{M_2023}\cite{Kingkar2022Cornell}\cite{PhysRevD.21.203}, in this potential describes the asymptotic behaviour of quarks at very short distances and arises from one-gluon exchange interactions. The linear confinement term, \(br\), was introduced by evaluating the Wilson loop in the long-distance regime, reflecting the confining nature of QCD. The parameter \(C\) represents a constant energy shift\cite{Kingkar2022Cornell}.  

The manuscript is organised as follows. In Section~II, we present the theoretical formalism, including the fitting procedure, the Matrix Numerov method, the variational method and the calculation of the meson properties. Section~III discusses the numerical results, and Section~IV presents the conclusions.

\section{Formalism}
\subsection{Parameter Estimation via Fitting of the Killingbeck and Cornell Potentials}
We applied linear regression using the OLS \cite{Press1989NumericalRecipes}\cite{goldberger1964econometric} method to fit the Killingbeck potential, $V_{\mathrm{Kill}}(r)$, to the Cornell potential, $V_{\mathrm{Cornell}}(r)$. We consider $V_{\text{Cornell}}(r)$ as the target function with known values of $\alpha_s$, $b$, and $C$. The objective of this formalism is to approximate this target function, $V_{\text{Cornell}}(r)$, using an alternative parameterisation, $V_{\text{kill}}(r)$, over a specific range $[r_{\text{min}}, r_{\text{max}}]$. The respective functional forms are given by
\begin{equation}
V_{\text{kill}}(r) = a' r^2 + b' r + \frac{c'}{r}
\label{eq:killingbeck_general}
\end{equation}
and
\begin{equation}
V_{\text{Cornell}}(r) = -\frac{4\alpha_s}{3r} + br + C.
\label{eq3}
\end{equation}

Clearly, $a'$, $b'$, and $c'$ are unknown parameters. To perform the fit numerically, the continuous radial domain is discretised into a sequence of $N$ equidistant points:
\begin{equation}
r_i = r_{\text{min}} + (i - 1)\Delta r.
\label{eq:discretized_r}
\end{equation}

The target function $V_{\text{Cornell}}(r)$ is evaluated at these $N$ discrete points to construct a column matrix $\mathbf{Y}$ such that
\begin{equation}
\mathbf{Y} = \begin{pmatrix}
V_{\text{Cornell}}(r_1) \\
V_{\text{Cornell}}(r_2) \\
\vdots \\
V_{\text{Cornell}}(r_N)
\end{pmatrix}.
\label{eq:matrix_Y}
\end{equation}
By invoking the principles of linear algebra, this system is compactly framed as an ordinary multivariate linear regression model:
\begin{equation}
\mathbf{Y} = \mathbf{X}\boldsymbol{\Theta} + \boldsymbol{\epsilon},
\label{eq:regression_model}
\end{equation}
where $\boldsymbol{\Theta} \in \mathbb{R}^{3 \times 1}$ denotes the target parameter vector containing the structural constants to be extracted, $\mathbf{X} \in \mathbb{R}^{N \times 3}$ is the configuration matrix containing the radial basis functions, and $\boldsymbol{\epsilon} \in \mathbb{R}^{N \times 1}$ represents the stochastic residual vector accounting for localized optimization mismatches at each grid node.
\begin{equation}
\Theta =
\begin{pmatrix}
a' \\
b' \\
c'
\end{pmatrix},
\quad
\boldsymbol{\epsilon} =
\begin{pmatrix}
\epsilon_1 \\
\epsilon_2 \\
\vdots \\
\epsilon_N
\end{pmatrix},
\quad
\mathbf{X} =
\begin{pmatrix}
r_1^2 & r_1 & \dfrac{1}{r_1} \\
r_2^2 & r_2 & \dfrac{1}{r_2} \\
\vdots & \vdots & \vdots \\
r_N^2 & r_N & \dfrac{1}{r_N}
\end{pmatrix}
\end{equation}

To evaluate the global error across the entire discretised grid of $N$ points, the individual residuals are squared and subsequently summed. Minimising this sum of squared residuals with respect to the parameter vector $\boldsymbol{\Theta}$ requires setting the gradient to zero, which yields the optimal parameter estimates.
\begin{equation}
\sum_{i=1}^{N} \epsilon_i^2
=
\sum_{i=1}^{N} \left[ V_1(r_i) - V_2(r_i) \right]^2
=
\|\mathbf{Y} - \mathbf{X}\boldsymbol{\Theta}\|^2
\end{equation}

\begin{equation}
\sum_{i=1}^{N} \epsilon_i^2
=
\mathbf{Y}^T \mathbf{Y}
- 2 \boldsymbol{\Theta}^T \mathbf{X}^T \mathbf{Y}
+ \boldsymbol{\Theta}^T \mathbf{X}^T \mathbf{X} \boldsymbol{\Theta}
\end{equation}
\begin{equation}
\nabla_{\boldsymbol{\Theta}} [\sum_{i=1}^{N} \epsilon_i^2 ]= 0
\end{equation}
\begin{equation}
-2\mathbf{X}^T \mathbf{Y} + 2\mathbf{X}^T \mathbf{X} \boldsymbol{\Theta} = 0
\end{equation}
\begin{equation}
\mathbf{X}^T \mathbf{X} \boldsymbol{\Theta} = \mathbf{X}^T \mathbf{Y}
\label{1122}
\end{equation}
The equation \eqref{1122} directly yields standard system of normal equations which gives the unique analytical solution for the optimal fitting coefficients.
\begin{equation}
{\boldsymbol{\Theta}} = (\mathbf{X}^T \mathbf{X})^{-1} \mathbf{X}^T \mathbf{Y}
\end{equation}

\subsection{Matrix Numerov Method}
We determined the ground-state energy, quark-spin interaction energy, and the spatial probability density at $N$ discrete points (including the ground state) for $B$, $B_s$, and $B_c$ mesons by numerically solving the Schrödinger equation via the Matrix Numerov method\cite{Pillai2012MatrixNumerov}. We considered the  Killingbeck potential in the Hamiltonian with the optimised parameters $a^\prime$, $b^\prime$, and $c^\prime$, and it is expressed as:

\begin{equation}
-\frac{1}{2\mu} \frac{d^2\psi(r)}{dr^2} + V(r)\psi(r) = E\psi(r)
\label{555}
\end{equation}
Where $\mu$ is the reduced mass of the meson and $E$ is the energy eigenvalue. To solve the differential equation in the domain $r \in (0, r_{\max}]$, we discretise the radial coordinate into $N$ equidistant grid points with a step size of $\Delta r = r_{\max}/N$. The grid coordinates are defined as $r_i = i\Delta r$, where $i = 1, 2, 3, \dots, N$. The local Numerov algorithm integrates differential equations of the form $\psi''(r) = f(r)\psi(r)$. For Eq.~\eqref{555}, $f(r)$ is defined as $-2\mu[E - V(r)]$. On this domain, discretized into $N$ grids evenly spaced by $\Delta r$, the integration formula becomes:
\begin{equation}
\psi_{i+1} = \frac{\psi_{i-1}\left(12 - (\Delta r)^2 f_{i-1}\right) - 2\psi_i\left(5(\Delta r)^2 f_i + 12\right)}{(\Delta r)^2 f_{i+1} - 12}
\end{equation}
Using this, equation \eqref{555} can be rearranged into the form:
\begin{equation}
\begin{aligned}
-\frac{1}{2\mu(\Delta r)^2} \Big( \psi_{i-1} &- 2\psi_i \\
&+ \Psi_{i+1} \Big) \\
&+ \frac{V_{i-1}\psi_{i-1} + 10V_i\psi_i + V_{i+1}\psi_{i+1}}{12} \\
&= E \frac{(\psi_{i-1} + 10\psi_i + \psi_{i+1})}{12}
\end{aligned}
\end{equation}
The terms 
$\frac{\psi_{i-1} - 2\psi_i + \psi_{i+1}}{(\Delta r)^2}$ 
and 
$\frac{\psi_{i-1} + 10\psi_i + \psi_{i+1}}{12}$ 
can be expressed as $N \times N$ matrices defined via the Kronecker delta as
\begin{align}
    A_{ij} &= \frac{1}{(\Delta r)^2} \left( \delta_{i, j-1} - 2\delta_{ij} + \delta_{i, j+1} \right), \\
    B_{ij} &= \frac{1}{12} \left( \delta_{i, j-1} + 10\delta_{ij} + \delta_{i, j+1} \right).
\end{align}

and Eq.~\eqref{555} becomes
\begin{equation}
    \hat{H}\Psi = -\frac{1}{2\mu} A \Psi + B V \Psi = E B \Psi.
\end{equation}
Multiplying by $B^{-1}$ yields
\begin{equation}
    \hat{H}\Psi = -\frac{1}{2\mu} B^{-1} A \Psi + V \Psi = E \Psi.
\end{equation}
Here, the term $-\frac{1}{2\mu} B^{-1} A$ represents the kinetic energy operator on an $N$-point grid. However, we must enforce Dirichlet boundary conditions to define the system as a bound state; thus, $\Psi_0 = \Psi_{N+1} = 0$.
The total wave function can be separated into its radial and angular components:
\begin{equation}
    \Psi_{n,l,m}(r,\theta,\Phi) = R_{n,l}(r) Y_l^m(\theta, \phi).
\end{equation}
For the ground state ($n=1, l=0, m=0$), the angular part simplifies to $Y_0^0(\theta, \phi) = 1/\sqrt{4\pi}$. Hence, the ground-state wave function becomes
\begin{equation}
    \Psi_{100} = \frac{R_{10}}{\sqrt{4\pi}}.
\end{equation}

\subsection{Variational Method:}
We consider a Gaussian trial wave-function as, 
\begin{equation}
    \Psi(r) = \left( \frac{\alpha^2}{\pi} \right)^{3/4} e^{-\frac{\alpha^2 r^2}{2}}
\end{equation}
where $\mu$ is the reduced mass of the quark-antiquark pairs, and $\alpha$ is the variational parameter.
The interaction potential used is the Killingbeck potential\cite{Killingbeck1978}. Now, the Hamiltonian of the \( Q\bar{Q} \) bound mesonic system is given by:
\begin{equation}
\hat{H} = -\frac{1}{2\mu} \left[ \frac{\partial^2}{\partial r^2} + \frac{2}{r} \frac{\partial}{\partial r} \right] + V_{Kill}(r)
\label{eq:3}
\end{equation}

Following the variational scheme, the ground state energy is:
\begin{equation}
\langle E \rangle = \langle H \rangle = \frac{3\hbar^2\alpha^2}{4\mu} + \frac{3a'}{2\alpha^2} + \frac{2b'}{\alpha\sqrt{\pi}} + \frac{2\alpha c'}{\sqrt{\pi}}
\label{eq:expectation}
\end{equation}
Now, minimising the energy, we can find the variational parameter for different bound states of meson.
\begin{equation}
    \frac{d}{d\alpha} \langle H \rangle = 0
    \label{min}
\end{equation}
Equation \eqref{min} yields the variational parameter $\alpha$, which, when substituted into the energy expression, provides the minimum energy.

\subsection{Masses of Heavy Flavored Mesons:}
The mass formula for pseudo-scalar mesons is given by\cite{Lucha1989Phenomenological}\cite{Hoque2020}\cite{Roy:2016txl}:

\begin{equation}
M_P = M_Q + M_{\bar{Q}} + \langle E \rangle + \langle H_{\text{SS}} \rangle
\label{eq:meson_mass}
\end{equation}

Here, $M_P$ is the mass of the pseudo-scalar meson, $M_{Q}$ and $M_{\bar{Q}}$ are the masses of the constituent quark and antiquark, respectively, $\langle E \rangle$ denotes the ground-state energy of the system, and $\langle H_{\mathrm{SS}} \rangle$ represents the energy shift due to mass splitting arising from quark spin interaction\cite{Lucha1989Phenomenological}, which is given by:

\begin{equation}
\langle H_{\text{SS}} \rangle = \frac{32 \pi \alpha_s}{9 M_Q M_{\bar{Q}}} 
\left( \vec{S}_Q \cdot \vec{S}_{\bar{Q}} \right) |\Psi_{100}(0)|^2
\end{equation}

Now, for pseudo-scalar mesons, the spin-spin interaction term becomes\cite{Griffiths1987ElementaryParticles}:
\[
\vec{S}_Q \cdot \vec{S}_{\bar{Q}} = -\frac{3}{4}
\] and so the mass formula becomes\cite{Pathak2011}\cite{Lahkar2019}
\begin{equation}
    M_p = M_Q + M_{\bar{Q}} + \langle E \rangle- \frac{8 \pi \alpha_s}{3 M_Q M_{\bar{Q}}} |\Psi_{100}(0)|^2
    \label{masseq}
\end{equation}

\subsection{Decay Constant}

Charged mesons can decay to a lepton-neutrino pair through annihilation via a virtual $W^{\pm}$ boson. The weak decays of mesons can be classified into three categories: leptonic, semileptonic, and non-leptonic decays \cite{dash2024purely}\cite{zhang2021comprehensive}\cite{Neubert2005EFTReview} according to the final products of the decay. Among these, leptonic decays are rarer but have a clear experimental signature due to the presence of a highly energetic lepton in the final state \cite{zhang2021comprehensive}, in which a charged lepton ($l = e, \mu, \tau$) and its corresponding neutrino are produced. Leptonic decays are characterised by a parameter called the decay constant, which quantifies the wave-function overlap between the quark and anti-quark. This is an important parameter for estimating the Cabibbo-Kobayashi-Maskawa (CKM) matrix elements\cite{pathak2026ckm} , as the decay rate depends on it. According to the Van-Royen-Weisskopf formula\cite{Pathak2011}\cite{royen1967hadron}, the decay constant of a pseudo-scalar meson is related to the wave-function at the origin as:

\begin{equation}
f_P \left( |\Psi_{100}(0)|^2, M_P \right) = \sqrt{\frac{12 |\Psi_{100}(0)|^2}{M_P}}
\label{dc1}
\end{equation}

Now introducing the QCD correction term\cite{lahkar2023meson}\cite{Lahkar2019}\cite{rai2008properties} it becomes,

\begin{equation}
f_P = \sqrt{\frac{12 |\Psi_{100}(0)|^2}{M_P}} \bar{C}
\label{dc2}
\end{equation}

where,
\begin{equation}
\bar{C} = \sqrt{1 - \frac{\alpha_s}{\pi} \left[ 2 - \frac{M_Q - M_{\bar{Q}}}{M_Q + M_{\bar{Q}}} \ln \left( \frac{M_Q}{M_{\bar{Q}}} \right) \right]}
\label{thfd}
\end{equation}
\subsection{Oscillation Frequency}

Neutral mesons, $B$ and $B_s$ exhibit the property of particle-antiparticle mixing\cite{Pathak2011}\cite{hulsbergen2013constraining} where particles make transitions into their antiparticle states and vice versa. The transitions $B_q^0 \leftrightarrow \bar{B}_q^0$ are due to the weak interaction \cite{Pathak2011} and caused by the difference between their weak and mass eigenstates\cite{ebert2003properties}\cite{buras2003relations}. This causes them to oscillate between these states\cite{Pathak2011}, and it is characterised by a mixing mass parameter, $\Delta M_B$, known as their frequency of oscillation.

\begin{equation}
\Delta M_B = \frac{G_F^2 m_t^2 M_{B_q} f_{B_q}^2}{8\pi} g(x_t) \eta_t |V_{tq}V_{tb}|^2 B,
\end{equation}

and,
\begin{equation}
q = d, s
\end{equation}

Where,
\begin{equation}
g(x_t) = \frac{1}{4} + \frac{9}{4(1 - x_t)} - \frac{3}{2(1 - x_t)^2} - \frac{3x_t^2}{2(1 - x_t)^3}
\end{equation}

Here, $G_F$ is the Fermi constant\cite{Pathak2011}, $m_t$ is the top quark mass\cite{Pathak2011}, $M_{B_q}$ and $f_{B_q}$ are the mass and decay constant of the $B_q$ meson, $g(x_t)$ is a loop function with $x_t = \frac{m_t^2}{M_W^2}$\cite{Pathak2011}, $\eta_t$ is a gluonic correction to oscillation\cite{Pathak2011}, $B$ is the bag parameter[19] and $V_{tq}, V_{tb}$ are CKM matrix elements\cite{Pathak2011}\cite{pathak2026ckm}.

\subsection{Isgur--Wise Function and Its Parameters}
The Isgur--Wise function is a universal form factor that describes the probability that the light-quark cloud remains completely undisturbed when the heavy quark undergoes a sudden change in velocity during a weak decay. In the heavy-quark limit ($m_Q \rightarrow \infty$), the spin and flavor degrees of freedom of the heavy quark decouple from the surrounding light-quark environment, leading to the emergence of an $SU(2N)$ spin--flavor symmetry. This enlarged symmetry reduces the complex set of independent hadronic form factors to a single universal function, known as the Isgur--Wise function (IWF)\cite{Hazarika2011arXiv}\cite{Das2016}. It is given by (writing as $\Psi_{100}(r)=\Psi(r)$)

\begin{equation}
\xi(Y)=\int_{0}^{\infty} 4\pi r^{2} |\Psi(r)|^{2}\cos(pr)\,dr
\label{eq:IWF}
\end{equation}

where,

\begin{equation}
p^{2}=2\mu^{2}(Y-1).
\label{eq:p}
\end{equation}

Here, $Y$ denotes the velocity transfer parameter, which characterises the recoil experienced during a weak semileptonic decay. At the zero-recoil point ($Y=1$), the initial and final mesons move with the same velocity, and the Isgur--Wise function is normalised to unity. As $Y$ increases beyond unity, the recoil of the final meson increases, resulting in a reduced overlap between the initial and final meson wave functions. Consequently, the probability of the light degrees of freedom remaining bound to the new heavy-quark core decreases. Expanding the cosine term in Eq.~(\ref{eq:IWF}) as a Taylor series, the Isgur--Wise function can be expressed as
\begin{equation}
\begin{aligned}
\xi(Y)=&\,1
-\left[
4\pi\mu^{2}
\int_{0}^{\infty}
r^{4}|\Psi(r)|^{2}\,dr
\right](Y-1) \\
&+\left[
\frac{2}{3}\pi\mu^{4}
\int_{0}^{\infty}
r^{6}|\Psi(r)|^{2}\,dr
\right](Y-1)^{2}
+\cdots .
\end{aligned}
\label{eq:IWFExpansion}
\end{equation}

Equation~(\ref{eq:IWFExpansion}) can be written in the standard form as\

\begin{equation}
\begin{aligned}
\xi(Y)=&\,1
-\rho^{2}(Y-1) \\
&+C(Y-1)^{2}
+\cdots .
\end{aligned}
\label{eq:IWFStandard}
\end{equation}
Where, 
\begin{equation}
\rho^{2}
=
-\left.
\frac{d\xi(Y)}{dY}
\right|_{Y=1},
\label{eq:slope}
\end{equation}

and

\begin{equation}
C
=
\frac{1}{2}
\left.
\frac{d^{2}\xi(Y)}{dY^{2}}
\right|_{Y=1},
\label{eq:curvature}
\end{equation}

At the zero-recoil point ($Y=1$), the Isgur--Wise function is normalised to unity, and its behaviour is governed primarily by the linear term in the expansion. The parameter $\rho^{2}$ denotes the slope of the Isgur--Wise function\cite{Das2016}\cite{Hazarika2011arXiv}, while $C$ represents its curvature\cite{Das2016}\cite{Hazarika2011arXiv}. Physically, the slope characterises the initial rate at which the overlap between the initial and final meson states decreases with increasing recoil. The curvature parameter quantifies the higher-order response of the light spectator-quark cloud to the recoil of the heavy quark and provides information about the dynamical deformation and stabilisation of the bound-state configuration during the decay process.

In the present matrix Numerov formalism, the ground-state wave function is written as

\begin{equation}
\Psi_{100}(r,\theta,\phi)
=
R_{10}(r)Y_{0}^{0}(\theta,\phi)
=
\frac{1}{\sqrt{4\pi}}\,R_{10}(r).
\label{eq:psi100}
\end{equation}

To express $R_{10}(r)$ in the standard Numerov formalism, we substitute

\begin{equation}
R_{10}(r)=\frac{\psi(r)}{r}.
\label{eq:Rpsi}
\end{equation}
and
\begin{equation}
 \Psi(r)=\frac{\psi(r)}{\sqrt{4\pi}r}   
\end{equation}
Using equations (\ref{eq:slope}) and (\ref{eq:curvature}), the integral expressions for the slope and curvature parameters are discretised on the radial grid. Consequently, the slope parameter becomes the following.

\begin{equation}
\rho^{2}
=
\mu^{2}
\sum_{i=1}^{N}
r_{i}^{2}
\left|\psi(r_{i})\right|^{2}
\Delta r,
\label{eq:slopeDiscrete}
\end{equation}

while the curvature parameter is given by

\begin{equation}
C
=
\frac{1}{6}\mu^{4}
\sum_{i=1}^{N}
r_{i}^{4}
\left|\psi(r_{i})\right|^{2}
\Delta r.
\label{eq:curvatureDiscrete}
\end{equation}

\section{Results}
\subsection{Estimation of the Unknown Killingbeck Potential parameter:}

We used the OLS \cite{Press1989NumericalRecipes} linear regression method to establish a functional correspondence between the Killingbeck potential and the Cornell potential, as discussed in the Formalism section. The values of $\alpha_s$ and $C$ were taken from Table \ref{parameters}\cite{Kingkar2022Cornell}. The parameter $b$ was fixed at $0.195$\cite{Choudhuryetal2026}, following our earlier work. The values of the Cornell potential parameters ($\alpha_S$, $C$) and Killingbeck parameters ($ a'$, $b'$, and $c'$) for $B$, $B_s$, and $B_c$ mesons are listed in Tables \ref{tab:merged_all_final} and \ref{tab:merged_all_finalvar} along with the corresponding physical properties that yield mass estimates in closest agreement with experimental results. The fitting is performed within the range of $0$ to $2.0 \text{ GeV}^{-1}$, which encompasses the charge radii of these mesons ($\sim 2.0 \text{ GeV}^{-1}$).
\begin{table}[h]
\centering
\caption{Allowed range of $\alpha_s$ and $C$ for different mesons under the constraints\cite{Kingkar2022Cornell}}
\label{parameters}
\begin{tabular}{lcc}
\toprule
Mesons & $\alpha_s$ & $C$ \\
\midrule

$B\;(\mu = 0.3157\ \text{GeV})$   & $0.580 - 0.629$ & $-0.998\ \text{to}\ -0.887$ \\
$B_s\;(\mu = 0.4238\ \text{GeV})$ & $0.450 - 0.582$ & $-0.997\ \text{to}\ -0.663$ \\
$B_c\;(\mu = 1.171\ \text{GeV})$  & $0.200 - 0.409$ & $-0.994\ \text{to}\ -0.197$ \\
\bottomrule
\end{tabular}
\end{table}
\begin{figure}[htbp]
    \centering
    \begin{subfigure}[b]{0.5\textwidth}
        \centering
        \includegraphics[width=\textwidth]{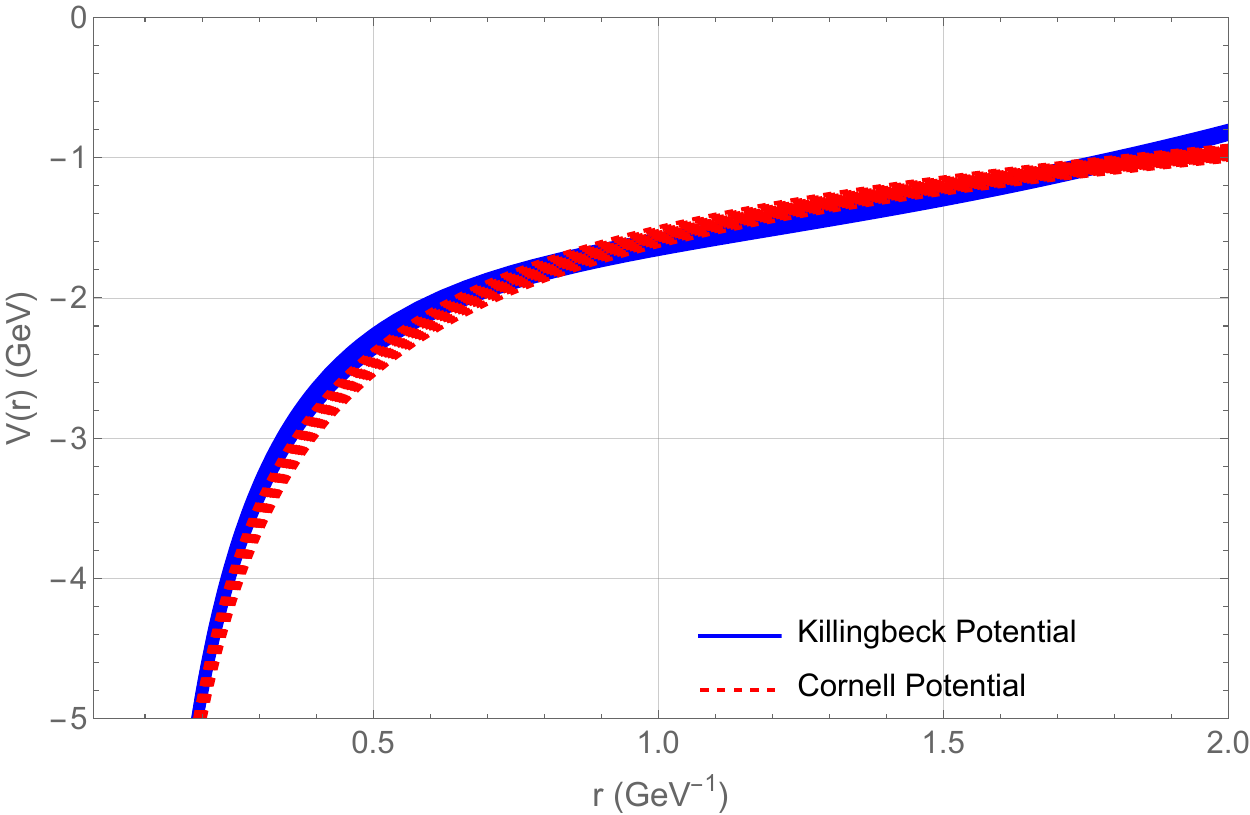}
        \caption{For $B$ meson}
        \label{fig:sub1}
    \end{subfigure}
    \hfill 
    \begin{subfigure}[b]{0.5\textwidth}
        \centering
        \includegraphics[width=\textwidth]{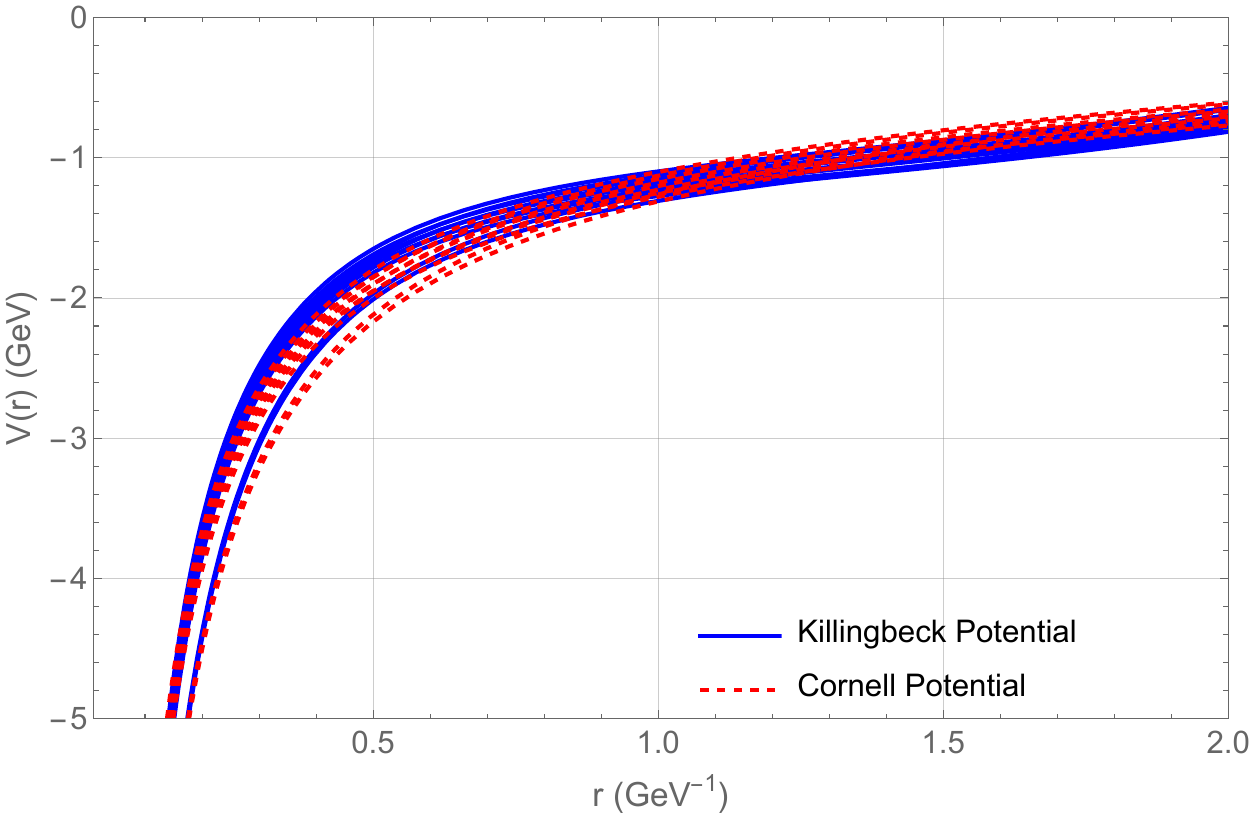}
        \caption{For $B_s$ meson}
        \label{fig:sub2}
    \end{subfigure}
    \hfill
    \begin{subfigure}[b]{0.5\textwidth}
        \centering
        \includegraphics[width=\textwidth]{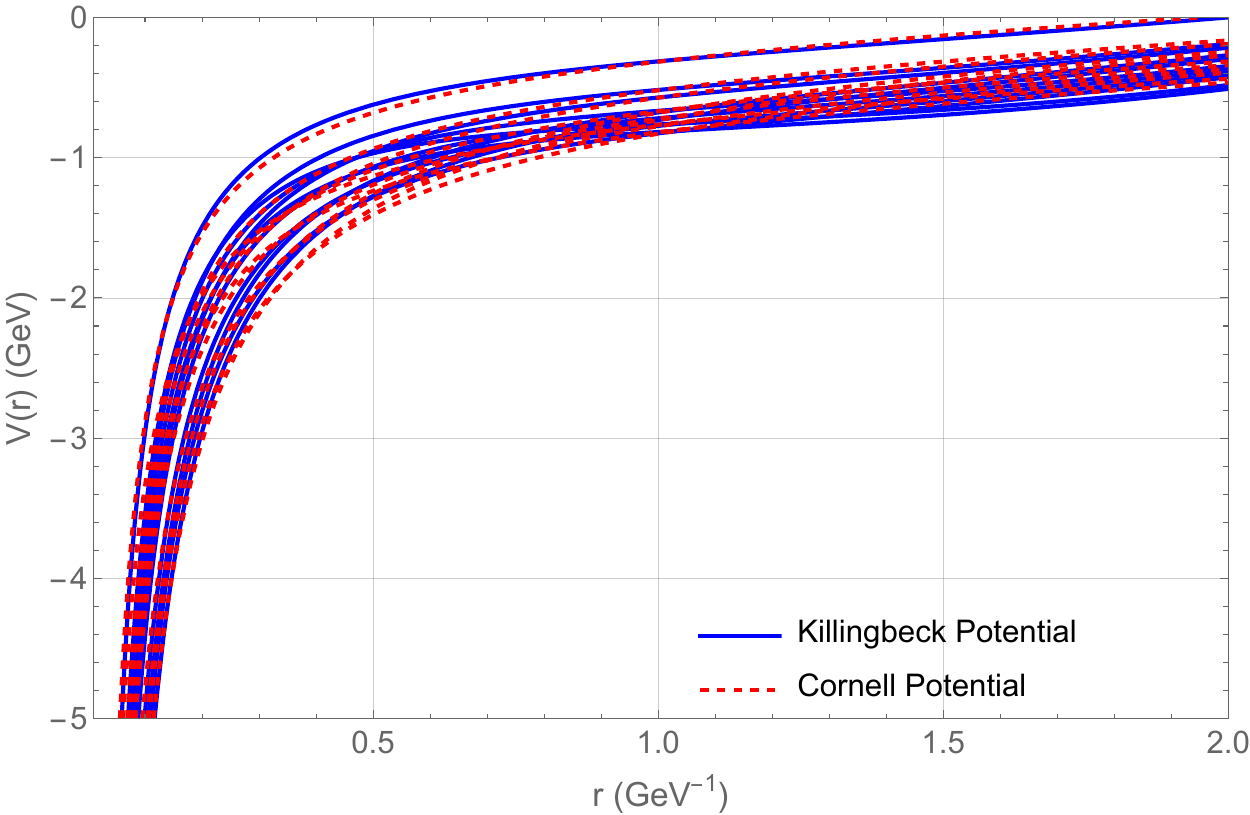}
        \caption{For $B_c$ meson}
        \label{fig:sub3}
    \end{subfigure}
    
    \caption{Least-squares fitting of the Killingbeck and Cornell potential for $B, B_s$ and $B_c$ meson.The solid and dashed lines represent the fitted theoretical curves obtained from the datasets in Table \ref{tab:merged_all_final} and Table \ref{tab:merged_all_finalvar}, respectively.}
    \label{fig:placeholder}
\end{figure}

\begin{figure}[htbp]
    \centering
    \begin{subfigure}[b]{0.5\textwidth}
        \centering
        \includegraphics[width=\textwidth]{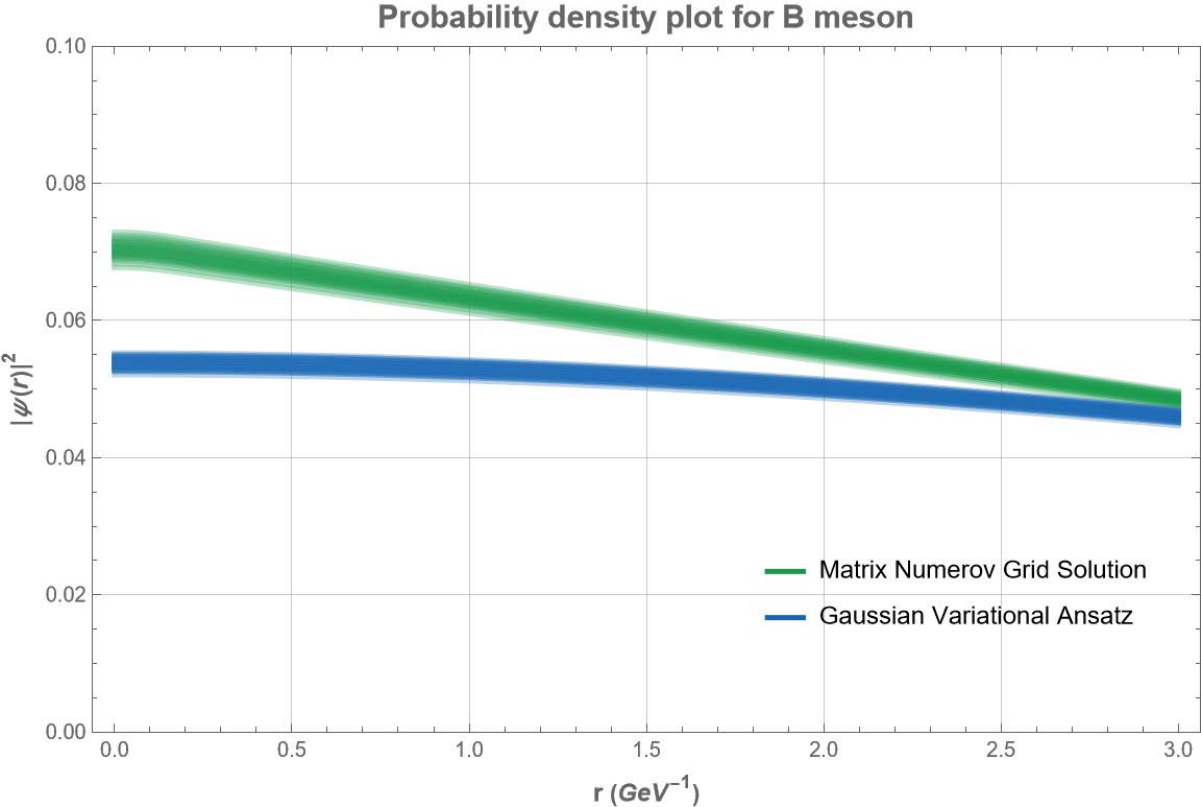}
        \caption{For $B$ meson}
        \label{fig:sub1}
    \end{subfigure}
    \hfill 
    \begin{subfigure}[b]{0.5\textwidth}
        \centering
        \includegraphics[width=\textwidth]{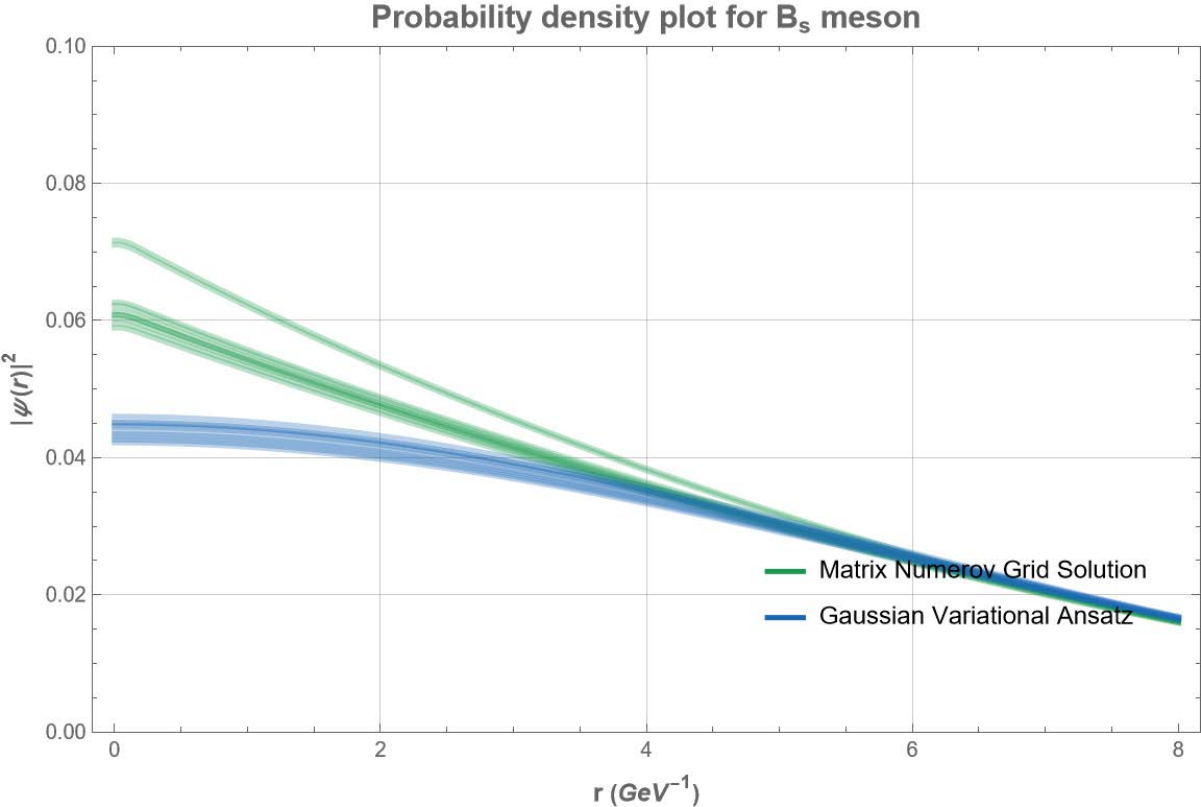}
        \caption{For $B_s$ meson}
        \label{fig:sub2}
    \end{subfigure}
    \hfill
    \begin{subfigure}[b]{0.5\textwidth}
        \centering
        \includegraphics[width=\textwidth]{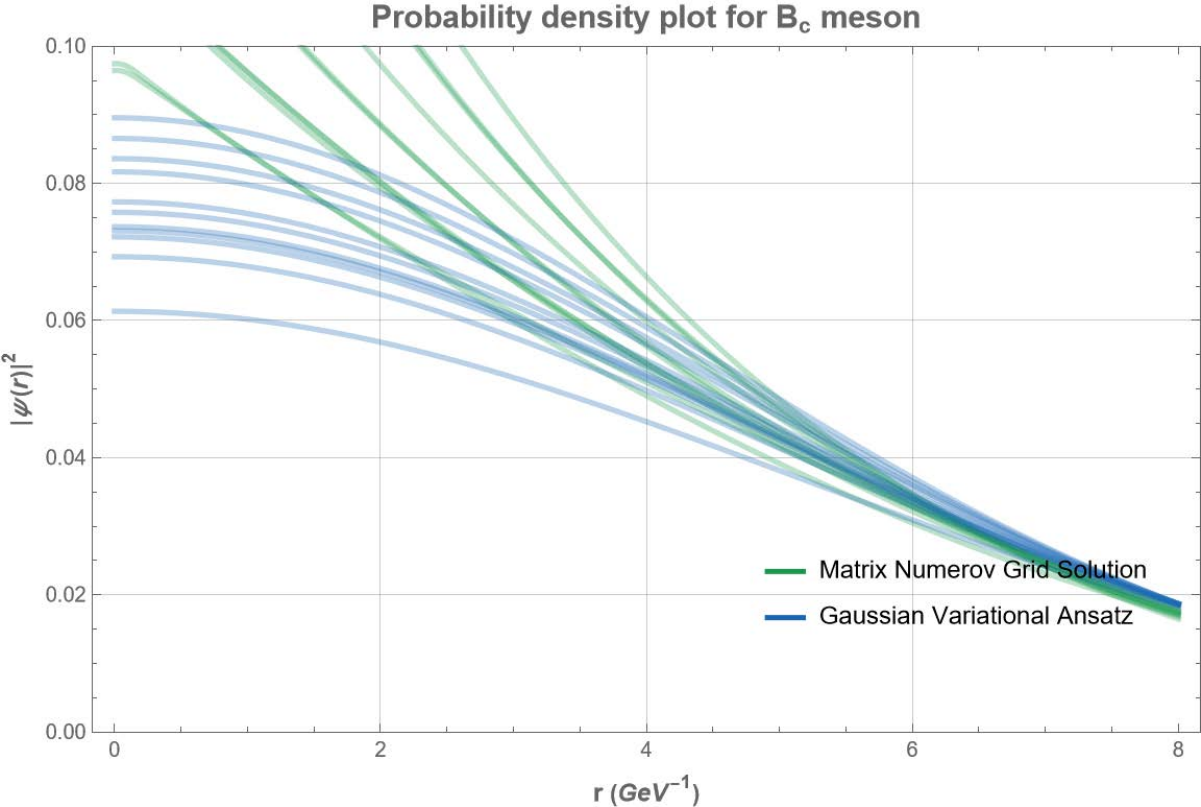}
        \caption{For $B_c$ meson}
        \label{fig:sub3}
    \end{subfigure}
    
    \caption{Radial probability distributions of $B$, $B_s$, and $B_c$ mesons calculated using the Matrix Numerov method and the variational method with a Gaussian trial wave function.}
    \label{fig:placeholder}
    \label{fig:three_figs}
\end{figure}

\subsection{\label{sec:level2}Mass}

As given in the formalism section, the pseudoscalar meson masses are 
calculated using the mass relation equation \eqref{masseq}, 
where $M_Q$, $M_{\bar{Q}}$, $\langle E_{\text{GS}} \rangle$, and $\langle H_{\text{SS}} 
\rangle$ are taken as input parameters and listed in Tables \ref{tab:merged_all_final} and \ref{tab:merged_all_finalvar}. The masses listed in table \ref{tab:merged_all_final} were calculated via the matrix-numerov method, whereas those in Table \ref{tab:merged_all_finalvar} were calculated via the variational method. The 
quark masses are taken to be the same as those in Refs.~\cite{Hoque2020, Das2016, 
Mutuk2018}, where the masses of the down ($M_d$), charm ($M_c$), strange ($M_s$), 
and bottom ($M_b$) quarks are set to 0.336, 1.550, 0.483, and 4.950 GeV, 
respectively. The coupling constants $\alpha_s$ in the expression for $\langle 
H_{\text{SS}} \rangle$ are also listed in the above-mentioned Tables. For the Numerov method, 
we set the range to $r_{\max} = 30\text{ GeV}^{-1}$ and discretized this domain 
into $N = 500$ points, which yields a step size of $\Delta r = r_{\max}/N = 0.06\text{ GeV}^{-1}$. 
We applied the condition $\psi_0 = \psi_{N+1} = 0$, which represents the Dirichlet boundary condition for our bounded system. However, for the ground state, the three-dimensional probability density function is given by $|\Psi(r)|^2 = \frac{1}{4\pi} |R(r)|^2$, where $R(r)$ is substituted as $\frac{\psi(r)}{r}$. Despite the Dirichlet boundary condition, the limiting value of $R(r)$ as $r \to 0$ is nonzero. On the other hand, Equations \eqref{eq:expectation} and \eqref{min} were used to determine the variational 
parameter $\alpha_{\min}$, which was subsequently applied in Equation~(24) to 
calculate the ground-state energy $\langle E \rangle$. Equation \eqref{masseq} was used to 
calculate the mass, whereas the quantities $\alpha_{\min}$, $\langle E \rangle$, $\langle H_{SS} \rangle$ and mass, along with the Killingbeck and Cornell parameters, are listed in Table \ref{tab:merged_all_finalvar}. Additionally, a detailed comparison 
with experimental results and other theoretical frameworks (such as the QCD sum rule, lattice QCD, and alternative potential models) is presented in Table \ref{tab:mass_comparison_all}.
\subsection{Decay Constant}
The decay constants obtained for each meson using the Matrix Numerov and variational methods are presented in Tables \ref{tab:merged_all_final} and \ref{tab:merged_all_finalvar}. It is calculated using equations \eqref{dc1}, \eqref{dc2}, and \eqref{thfd}, where our calculated meson masses are used as input parameters. The correction factor $\bar{C}$ is computed as $0.9$ using equation \eqref{thfd}. Our calculated decay constants are compared with experimental values and theoretical predictions from other models, such as QCD sum rules, lattice QCD, and other potential models in the literature (See table \ref{tab:decay_constant_comparison_all}).

\subsection{Oscillation Frequency}

As explained in the formalism, the oscillation frequencies for $B$ and $B_s$ mesons are calculated using Equations (16) and (17). The calculation incorporates the Fermi coupling constant $G_F = 1.166 \times 10^{-5}\text{ GeV}^{-2}$, the parameter $x_t = m_t^2/m_W^2 = 4.730$, and the loop function $g(x_t) = 0.189$. The other parameters involved in these expressions include $m_t = 174\text{ GeV}$\cite{lahkar2023meson}\cite{Pathak2011}, $m_W = 80.403\text{ GeV}$\cite{lahkar2023meson}\cite{Pathak2011}, the Bag parameter $B = 1.34$\cite{Pathak2011}, and the elements of the CKM matrix: $|V_{td}| = 0.0074$\cite{Pathak2011}, $|V_{ts}| = 0.04$\cite{Pathak2011}\cite{pathak2026ckm}, and $|V_{tb}| = 1$. The QCD correction factor $\eta_t$ is taken as $0.55$\cite{lahkar2023meson}\cite{Lahkar2019}\cite{Pathak2011}. Our calculated values and the comparison of these values with the experimental values and theoretically predicted values of the other potential model are listed in Table \ref{tab:oscillation_frequency_comparison_all}.

\subsection{Isgur-Wise Parameters: Slope $\rho^2$ and Curvature $C$}
\label{subsec:isgur_wise}

The Isgur-Wise parameters: the slope ($\rho^2$) and curvature ($C$), are explicitly defined via equations \eqref{eq:IWFExpansion},\eqref{eq:IWFStandard},\eqref{eq:slope},\eqref{eq:curvature}. In the context of the Numerov numerical framework, the wave function is transformed using the radial substitution $\Psi(r) = \frac{1}{\sqrt{4\pi}} \frac{\psi(r)}{r}$, thereby reducing the governing integro-differential equations to the discretised representations given in equations \eqref{eq:slopeDiscrete} and \eqref{eq:curvatureDiscrete}. Conversely, the 
variational approach directly utilises the continuous integral formulation detailed in equations \eqref{eq:IWFExpansion} and \eqref{eq:IWFStandard}. For comparative analysis, the physical observables computed via both the variational and Numerov methods 
-namely the mass spectrum, decay constants, and mixing oscillation frequencies 
-are consolidated alongside the Isgur-Wise parameters in Tables  \ref{tab:merged_all_final} and \ref{tab:merged_all_finalvar}.
\begin{figure}[htbp]
    \centering
    \begin{subfigure}[b]{0.5\textwidth}
        \centering
        \includegraphics[width=\textwidth]{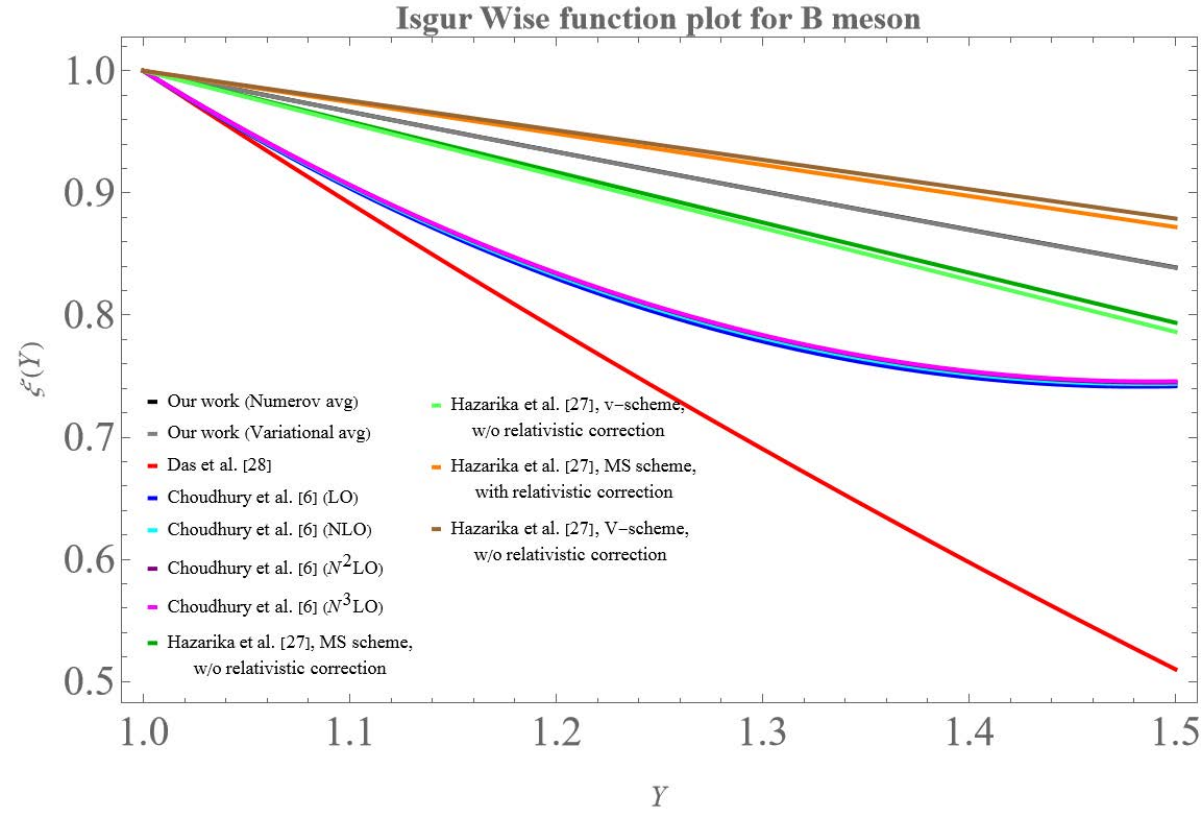}
        \caption{For $B$ meson}
        \label{fig:sub1}
    \end{subfigure}
    \hfill 
    \begin{subfigure}[b]{0.5\textwidth}
        \centering
        \includegraphics[width=\textwidth]{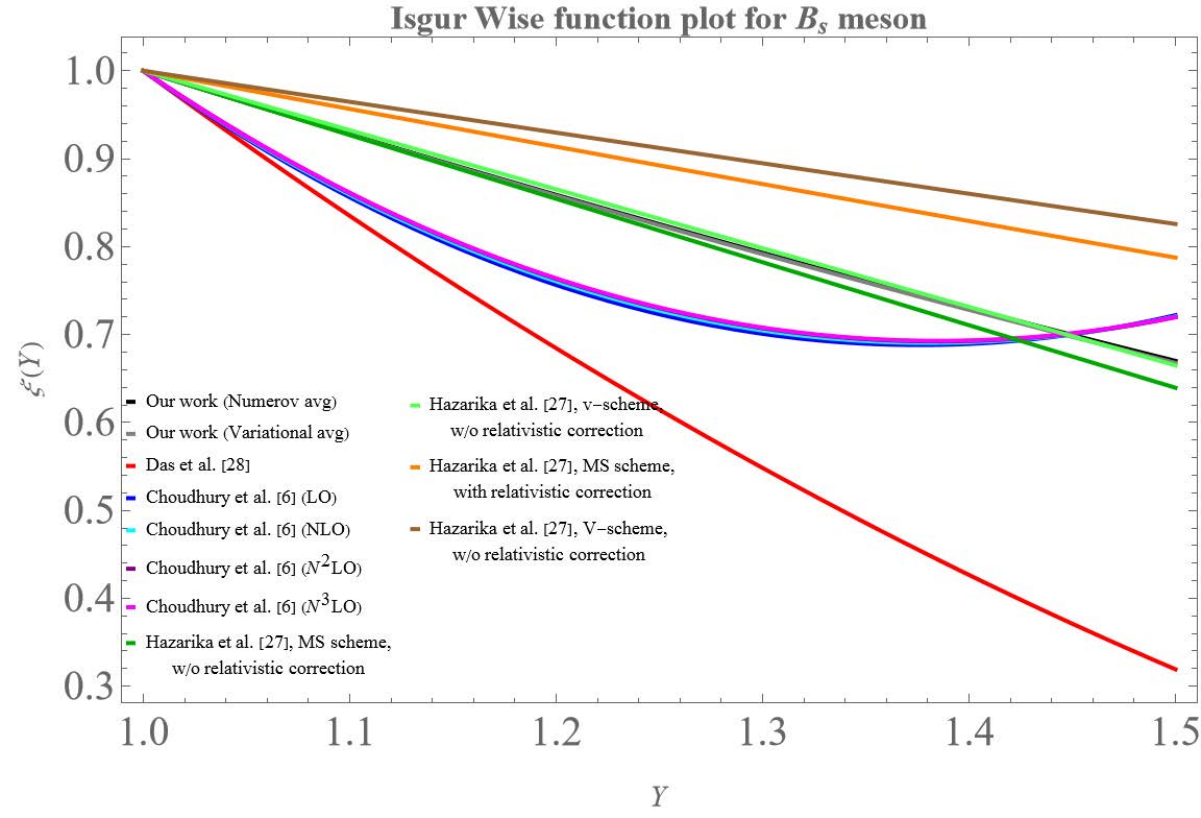}
        \caption{For $B_s$ meson}
        \label{fig:sub2}
    \end{subfigure}
    \hfill
    \begin{subfigure}[b]{0.5\textwidth}
        \centering
        \includegraphics[width=\textwidth]{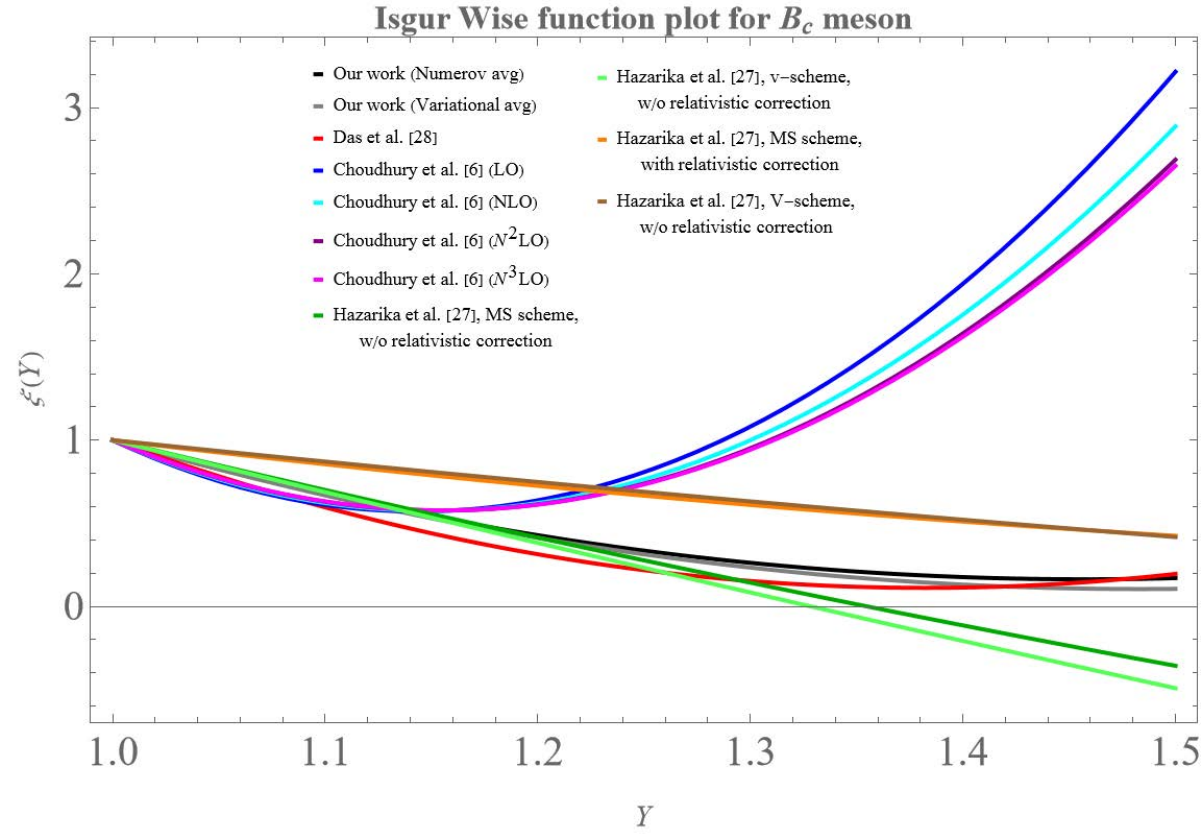}
        \caption{For $B_c$ meson}
        \label{fig:sub3}
    \end{subfigure}
    
    \caption{Variation of $\xi(Y)$ with the recoil parameter $Y = v \cdot v'$ across the heavy meson sector, including a comparison of the parameters $\rho^2$ and $C$ with values from other works (From tables~\ref{tab:slope_comparison_all} and \ref{tab:curve_comparison_all}).}
    \label{fig:placeholder}
    \label{fig:three_figs}
\end{figure}

\begin{table*}[t]
\centering
\footnotesize
\setlength{\tabcolsep}{1pt}
\caption{Numerical Results for the Meson Properties and Isgur--Wise
Function Parameters Using the Matrix Numerov Method.}
\label{tab:merged_all_final}

\begin{ruledtabular}
\begin{tabular}{ccccccccccccc}
$\alpha_s$
& \shortstack{$C_{\rm pot}$ \\ (GeV)}
& \shortstack{$a'$ \\ (GeV$^3$)}
& \shortstack{$b'$ \\ (GeV$^2$)}
& $c'$
& $|\Psi(0)|^2$
& \shortstack{$\langle E\rangle_{\rm gs}$ \\ (GeV)}
& \shortstack{$\langle H_{\rm SS}\rangle$ \\ (GeV)}
& \shortstack{Mass \\ (GeV)}
& \shortstack{$f_P$ \\ (GeV)}
& \shortstack{$\Delta m_B$ \\ (${\rm ps}^{-1}$)}
& $\rho^2$
& $C$ \\
\hline

\multicolumn{13}{c}{$B$ meson} \\
\hline

\shortstack{$0.600$ \\ $\pm0.014$} & \shortstack{$-0.958$ \\ $\pm0.028$} & \shortstack{$0.521$ \\ $\pm0.017$} & \shortstack{$-1.222$ \\ $\pm0.042$} & \shortstack{$-0.890$ \\ $\pm0.003$} & \shortstack{$0.072$ \\ $\pm0.001$} & \shortstack{$0.045$ \\ $\pm0.026$} & \shortstack{$0.073$ \\ $\pm0.003$} & \shortstack{$5.258$ \\ $\pm0.028$} & \shortstack{$0.365$ \\ $\pm0.004$} & \shortstack{$0.891$ \\ $\pm0.014$} & \shortstack{$0.338$ \\ $\pm0.004$} & \shortstack{$0.032$ \\ 
$\pm6.8\times10^{-4}$} \\

\hline
\multicolumn{13}{c}{$B_s$ meson} \\
\hline

\shortstack{$0.482$ \\ $\pm0.047$} & \shortstack{$-0.762$ \\ $\pm0.058$} & \shortstack{$0.281$ \\ $\pm0.020$} & \shortstack{$-0.743$ \\ $\pm0.068$} & \shortstack{$-0.743$ \\ $\pm0.058$} & \shortstack{$0.017$ \\ $\pm0.002$} & \shortstack{$-0.124$ \\ $\pm0.053$} & \shortstack{$0.010$ \\ $\pm0.002$} & \shortstack{$5.300$ \\ $\pm0.057$} & \shortstack{$0.176$ \\ $\pm0.009$} & \shortstack{$8.982$ \\ $\pm0.767$} & \shortstack{$0.736$ \\ $\pm0.024$} & \shortstack{$0.152$ \\ 
$\pm0.006$} \\

\hline
\multicolumn{13}{c}{$B_c$ meson} \\
\hline

\shortstack{$0.284$ \\ $\pm0.065$} & \shortstack{$-0.490$ \\ $\pm0.132$} & \shortstack{$0.184$ \\ $\pm0.050$} & \shortstack{$-0.419$ \\ $\pm0.166$} & \shortstack{$-0.436$ \\ $\pm0.085$} & \shortstack{$0.137$ \\ $\pm0.032$} & \shortstack{$-0.156$ \\ $\pm0.138$} & \shortstack{$0.010$ \\ $\pm0.002$} & \shortstack{$6.301$ \\ $\pm0.089$} & \shortstack{$0.456$ \\ $\pm0.053$} & $\cdot$ & \shortstack{$3.660$ \\ $\pm0.283$} & \shortstack{$3.897$ \\ 
$\pm0.584$} \\
\end{tabular}
\end{ruledtabular}
\end{table*}

\begin{table*}[t]
\centering
\footnotesize
\setlength{\tabcolsep}{1pt}
\caption{Numerical Results for the Meson Properties and Isgur--Wise
Function Parameters Using the Variational Method.}
\label{tab:merged_all_finalvar}

\begin{ruledtabular}
\begin{tabular}{cccccccccccccc}
$\alpha_s$
& \shortstack{$C_{\rm pot}$ \\ (GeV)}
& \shortstack{$a'$ \\ (GeV$^3$)}
& \shortstack{$b'$ \\ (GeV$^2$)}
& $c'$
& $|\Psi(0)|^2$
& $\alpha_{min}$
& \shortstack{$\langle E\rangle_{\rm gs}$ \\ (GeV)}
& \shortstack{$\langle H_{\rm SS}\rangle$ \\ (GeV)}
& \shortstack{Mass \\ (GeV)}
& \shortstack{$f_P$ \\ (GeV)}
& \shortstack{$\Delta m_B$ \\ (${\rm ps}^{-1}$)}
& $\rho^2$
& $C$ \\
\hline

\multicolumn{14}{c}{$B$ meson} \\
\hline

\shortstack{$0.600$ \\ $\pm0.014$} & \shortstack{$-0.958$ \\ $\pm0.028$} & \shortstack{$0.521$ \\ $\pm0.017$} & \shortstack{$-1.222$ \\ $\pm0.042$} & \shortstack{$-0.890$ \\ $\pm9.0\times10^{-3}$} & \shortstack{$0.054$ \\ $\pm0.030$} & \shortstack{$0.670$ \\ $\pm0.004$} & \shortstack{$0.053$ \\ $\pm0.030$} & \shortstack{$0.054$ \\ $\pm0.002$} & \shortstack{$5.286$ \\ $\pm0.031$} & \shortstack{$0.314$ \\ $\pm0.003$} & \shortstack{$0.983$ \\ $\pm0.015$} & \shortstack{$0.338$ \\ $\pm0.004$} & \shortstack{$0.032$ \\ $\pm5.6\times10^{-4}$} \\

\hline
\multicolumn{14}{c}{$B_s$ meson} \\
\hline

\shortstack{$0.482$ \\ $\pm0.047$} & \shortstack{$-0.762$ \\ $\pm0.058$} & \shortstack{$0.281$ \\ $\pm0.020$} & \shortstack{$-0.743$ \\ $\pm0.068$} & \shortstack{$-0.743$ \\ $\pm0.058$} & \shortstack{$0.044$ \\ $\pm0.001$} & \shortstack{$0.626$ \\ $\pm0.006$} & \shortstack{$-0.121$ \\ $\pm0.055$} & \shortstack{$0.025$ \\ $\pm0.003$} & \shortstack{$5.287$ \\ $\pm0.057$} & \shortstack{$0.285$ \\ $\pm0.006$} & \shortstack{$23.510$ \\ $\pm0.676$} & \shortstack{$0.741$ \\ $\pm0.014$} & \shortstack{$0.153$ \\ $\pm0.006$} \\

\hline
\multicolumn{14}{c}{$B_c$ meson} \\
\hline

\shortstack{$0.284$ \\ $\pm0.065$} & \shortstack{$-0.490$ \\ $\pm0.132$} & \shortstack{$0.184$ \\ $\pm0.050$} & \shortstack{$-0.419$ \\ $\pm0.166$} & \shortstack{$-0.436$ \\ $\pm0.085$} & \shortstack{$0.076$ \\ $\pm0.008$} & \shortstack{$0.750$ \\ $\pm0.027$} & \shortstack{$-0.152$ \\ $\pm0.152$} & \shortstack{$0.007$ \\ $\pm0.003$} & \shortstack{$6.341$ \\ $\pm0.146$} & \shortstack{$0.343$ \\ $\pm0.021$} & $\cdot$ & \shortstack{$3.726$ \\ $\pm0.275$} & \shortstack{$3.703$ \\ $\pm0.544$} \\
\end{tabular}
\end{ruledtabular}
\end{table*}

\begin{table*}[t]
\centering
\caption{Comprehensive Comparison of Calculated and Experimental Ground-State Masses for $B$, $B_s$, and $B_c$ Mesons.}
\label{tab:mass_comparison_all}
\begin{ruledtabular}
\begin{tabular}{lccc}
Property / Metric & $B$ Meson (GeV) & $B_s$ Meson (GeV) & $B_c$ Meson (GeV) \\
\hline
Calculated Mass (Numerov Average)      & $5.258 \pm 0.028$ & $5.300 \pm 0.057$ & $6.301 \pm 0.093$ \\
Calculated Mass (Variational Average)  & $5.286 \pm 0.031$ & $5.287 \pm 0.057$ & $6.341 \pm 0.146$ \\
Closest Obtained Value (Numerov)       & $5.280$           & $5.346$           & $6.280$           \\
Closest Obtained Value (Variational)   & $5.280$           & $5.358$           & $6.246$           \\
Experimental Mass (PDG)\cite{PDG2024}  & $5.279\pm7\times10^{-5}$ & $5.366\pm10^{-4}$ & $6.274\pm3.2\times10^{-4}$ \\
LHCB2019,LHCB2020\cite{LHCb2020BcMass}\cite{LHCb2020BMass}\cite{LHCb2019BsMass} & $5.279\pm10^{-4}$ & $5.366\pm1.9\times10^{-4}$ & $6.274\pm2.7\times10^{-4}$ \\
F. Abe et al.\cite{CDF1998BcDiscovery} & $\cdot$           & $\cdot$           & $6.400\pm0.390$   \\
Hoque et al.\cite{Hoque2023}           & $5.260$           & $5.390$           & $6.439$           \\
Lahkar et al.\cite{Lahkar2019}         & $5.350^a,5.110^b,5.140^c$ & $5.480^a,5.400^b,5.372^c$ & $6.400^a,6.380^b,6.500^c$ \\
Aziz et al.\cite{Lahkar2019}          & $5.241$           & $5.329$           & $\cdot$           \\
Pathak et al.\cite{Pathak2011}        & $5.273^f,5.256^g$ & $5.370^f,5.349^g$ & $\cdot$           \\
\end{tabular}
\end{ruledtabular}
\begin{tablenotes}
\item a) Cornell potential with Gaussian$^a$, Coulomb$^b$ and Airy$^c$ wavefunctions.
\item b) For $^f n_f=3$, $^g n_f=4$
\end{tablenotes}
\end{table*}
\begin{table*}[t]
\centering
\caption{Comprehensive Comparison of Calculated and Experimental/Lattice Decay Constants ($f_P$) for $B$, $B_s$, and $B_c$ Mesons.}
\label{tab:decay_constant_comparison_all}
\begin{ruledtabular}
\begin{tabular}{lccc}
Property / Metric & $B$ Meson (GeV) & $B_s$ Meson (GeV) & $B_c$ Meson (GeV) \\
\hline
Calculated $f_P$ (Numerov Average)      & $0.365 \pm 0.004$ & $0.176 \pm 0.009$ & $0.456 \pm 0.053$ \\
Calculated $f_P$ (Variational Average)  & $0.314 \pm 0.003$ & $0.285 \pm 0.006$ & $0.343 \pm 0.021$ \\
Experimental value\cite{PhysRevD.78.052003}\cite{201455} & $0.198\pm0.014$ & $0.237\pm0.017$ & $\cdot$ \\
Lattice QCD\cite{Dowdall_2012}\cite{PhysRevD.86.034506}\cite{2014231} & $0.218\pm0.003$ & $0.228\pm0.010$ & $\cdot$ \\
QCD Sum Rule\cite{Wang2015}\cite{Lucha_2011} & $0.193\pm0.012$ & $0.232\pm0.018$ & $\cdot$ \\
FLAG Review 2024\cite{FLAG:2024Review} & $0.190$ & $0.230$ & $\cdot$ \\
Hoque et al.\cite{Hoque2023} & $0.213$ & $0.298$ & $0.579$ \\
Mutuk, 2018\cite{Mutuk2018} & $0.204$ & $\cdot$ & $\cdot$ \\
Lahkar et al.\cite{Lahkar2019} & $0.198^a,0.264^b,0.764^c$ & $0.207^a,0.238^b,0.627^c$ & $0.563^a,0.590^b,0.333^c$ \\
Pathak et al.\cite{Pathak2011} & $0.213^f,0.246^g$ & $0.265^f,0.311^g$ & $\cdot$ \\
\end{tabular}
\end{ruledtabular}
\begin{tablenotes}
\item a) Cornell potential with Gaussian$^a$, Coulomb$^b$ and Airy$^c$ wavefunctions.
\item b) For $^f n_f=3$, $^g n_f=4$
\end{tablenotes}
\end{table*}

\begin{table*}[t]
\centering
\caption{Comprehensive Comparison of Calculated and Experimental Mass Oscillation Frequencies ($\Delta m_B$) for $B$, $B_s$, and $B_c$ Mesons.}
\label{tab:oscillation_frequency_comparison_all}
\begin{ruledtabular}
\begin{tabular}{lccc}
Property / Metric & $B$ Meson (${\rm ps}^{-1}$) & $B_s$ Meson (${\rm ps}^{-1}$) & $B_c$ Meson (${\rm ps}^{-1}$) \\
\hline
Calculated $\Delta m_B$ (Numerov Average)      & $0.891 \pm 0.014$ & $8.982 \pm 0.809$  & $\cdot$ \\
Calculated $\Delta m_B$ (Variational Average)  & $0.982 \pm 0.015$ & $23.511 \pm 0.676$ & $\cdot$ \\
Experimental Value\cite{LHCb2013BsOscillation}\cite{LHCb2016B0Oscillation} & $0.507$ & $17.765$ & $\cdot$ \\
QCD Sum Rule\cite{Gay2000BMixing}              & $0.480$ & $>14.600$ & $\cdot$ \\
Lattice QCD\cite{Bazavov2016BmixingLattice}    & $0.630$ & $19.6$ & $\cdot$ \\
Lahkar et al.\cite{Lahkar2019}                 & $0.450^a,0.780^b,0.270^c$ & $16^a,60^b,9.3^c$ & $\cdot$ \\
Pathak et al.\cite{Pathak2011}                 & $0.550^f,0.740^g$ & $17.34^f,23.88^g$ & $\cdot$ \\
\end{tabular}
\end{ruledtabular}
\begin{tablenotes}
\item a) Cornell potential with Gaussian$^a$, Coulomb$^b$ and Airy$^c$ wavefunctions.
\item b) For $^f n_f=3$, $^g n_f=4$
\end{tablenotes}
\end{table*}

\begin{table*}[t]
\centering
\caption{Comprehensive Comparison of Calculated and Reference Isgur--Wise Slope Parameters ($\rho^2$) for $B$, $B_s$, and $B_c$ Mesons.}
\label{tab:slope_comparison_all}
\begin{ruledtabular}
\begin{tabular}{lccc}
Property / Metric & $B$ Meson & $B_s$ Meson & $B_c$ Meson \\
\hline
Calculated $\rho^2$ (Numerov Average)      & $0.338 \pm 0.004$ & $0.736 \pm 0.015$ & $3.659 \pm 0.283$ \\
Calculated $\rho^2$ (Variational Average)  & $0.339 \pm 0.003$ & $0.741 \pm 0.014$ & $3.703 \pm 0.272$ \\
\hline
Choudhury et al.\cite{Choudhuryetal2026} & 
  \begin{tabular}{@{}ll@{}} 
    LO:    & $1.072 \pm 0.124^a$ \\ 
    NLO:   & $1.054 \pm 0.121^b$ \\ 
    N$^2$LO: & $1.042 \pm 0.119^c$ \\ 
    N$^3$LO: & $1.040 \pm 0.118^d$ 
  \end{tabular} & 
  \begin{tabular}{c} $1.659\pm0.187$ \\ $1.623\pm0.181$\\ $1.599\pm0.176$ \\ $1.594\pm0.175$ \end{tabular} & 
  \begin{tabular}{c} $5.970\pm0.558$ \\ $5.666\pm0.508$ \\ $5.468\pm0.477$ \\ $5.430\pm0.471$ \end{tabular} \\
\hline
Das et al.\cite{Das2016}      & $1.110$           & $1.722$           & $4.646$           \\
Bjorken lower limit\cite{oliver2014isgur}        & $>0.25$           & $>0.25$           & $\cdot$           \\
Uraltsev's lower limit\cite{oliver2014isgur}        & $>0.75$           & $>0.75$           &$\cdot$\\
Le Yaouanc et al.\cite{Yaouanc2003}        & $>0.75$           & $>0.75$           & $\cdot$           \\
\hline
Hazarika et al.\cite{Hazarika2011arXiv}      & 
  \begin{tabular}{@{}c@{}} 
    $0.416^{a_1}$ \\ $0.431^{a_2}$ \\ $0.258^{b_1}$ \\ $0.244^{b_2}$ 
  \end{tabular} & 
  \begin{tabular}{@{}c@{}} 
    $0.733^{a_1}$ \\ $0.677^{a_2}$ \\ $0.436^{b_1}$ \\ $0.354^{b_2}$ 
  \end{tabular} & 
  \begin{tabular}{@{}c@{}} 
    $3.070^{a_1}$ \\ $3.160^{a_2}$ \\ $1.520^{b_1}$ \\ $1.320^{b_2}$ 
  \end{tabular} \\
\hline
\end{tabular}
\end{ruledtabular}
\begin{tablenotes}
    \small
    \item[$^{a,b,c,d}$] 0, 1, 2, and 3-loop corrections to the running coupling constant $\alpha_s$, respectively.
    \item[$^{a_1}$] $\overline{\text{MS}}$ scheme results without relativistic correction$^{a_1}$.
    \item[$^{a_2}$] $V$-scheme results without relativistic correction$^{a_2}$.
    \item[$^{b_1}$] $\overline{\text{MS}}$ scheme results with relativistic correction$^{b_1}$
    \item[$^{b_2}$] $V$-scheme results with relativistic correction$^{b_2}$.
\end{tablenotes}
\end{table*}
\begin{table*}[t]
\centering
\caption{Comprehensive Comparison of Calculated and Reference Isgur--Wise Curvature ($C$) Parameters for $B$, $B_s$, and $B_c$ Mesons.}
\label{tab:curve_comparison_all}
\begin{ruledtabular}
\begin{tabular}{lccc}
Property / Metric & $B$ Meson & $B_s$ Meson & $B_c$ Meson \\
\hline
Calculated $C$ (Numerov Average)       & $0.032 \pm 6.6\times10^{-4}$ & $0.152 \pm 0.006$ & $4.0 \pm 0.548$ \\
Calculated $C$ (Variational Average)   & $0.033 \pm 5.6\times10^{-3}$ & $0.153 \pm 0.006$ & $3.828 \pm 0.544$ \\
\hline
Choudhury et al.\cite{Choudhuryetal2026} & 
  \begin{tabular}{@{}ll@{}} 
    LO:    & $1.111 \pm 0.196^a$ \\ 
    NLO:   & $1.083 \pm 0.188^b$ \\ 
    N$^2$LO: & $1.064 \pm 0.183^c$ \\ 
    N$^3$LO: & $1.061 \pm 0.182^d$ 
  \end{tabular} & 
  \begin{tabular}{c} $2.206\pm0.404$ \\ $2.128\pm0.383$\\ $2.077\pm0.370$ \\ $2.068\pm0.367$ \end{tabular} & 
  \begin{tabular}{c} $20.803\pm3.609$ \\ $18.882\pm3.133$\\ $17.683\pm2.845$ \\ $17.460\pm2.792$ \end{tabular} \\
\hline
Das et al.\cite{Das2016}       & $0.260$           & $0.721$           & $6.074$           \\
Le Yaouanc et al.\cite{Yaouanc2003}        & $0.47$            & $0.47$            & $\cdot$           \\
\hline
Hazarika et al.\cite{Hazarika2011arXiv}      & 
  \begin{tabular}{@{}c@{}} 
    ${}^{a_1}6.87\times10^{-3}$ \\ ${}^{a_2}7.11\times10^{-3}$ \\ ${}^{b_1}4.26\times10^{-3}$ \\ ${}^{b_2}4.0\times10^{-3}$ 
  \end{tabular} & 
  \begin{tabular}{@{}c@{}} 
    $0.024^{a_1}$ \\ $0.014^{a_2}$ \\ $0.022^{b_1}$ \\ $0.011^{b_2}$ 
  \end{tabular} & 
  \begin{tabular}{@{}c@{}} 
    $0.712^{a_1}$ \\ $0.353^{a_2}$ \\ $0.740^{b_1}$ \\ $0.310^{b_2}$ 
  \end{tabular} \\
\hline
\end{tabular}
\end{ruledtabular}
\begin{tablenotes}
    \small
    \item[$^{a,b,c,d}$] 0, 1, 2, and 3-loop corrections to the running coupling constant $\alpha_s$, respectively.\cite{Choudhuryetal2026}
    \item[$^{a_1}$] $\overline{\text{MS}}$ scheme results without relativistic correction.$^{a_1}$
    \item[$^{a_2}$] $V$-scheme results without relativistic correction.$^{a_2}$
    \item[$^{b_1}$] $\overline{\text{MS}}$ scheme results with relativistic correction.$^{b_1}$
    \item[$^{b_2}$] $V$-scheme results with relativistic correction$^{b_2}$.
\end{tablenotes}
\end{table*}
\section{Conclusion}
In this study, we use the matrix Numerov method and the quantum mechanical variational method to investigate the spectroscopic and structural properties of the $B$, $B_s$, and $B_c$ mesons. Across all three meson systems, both methods consistently yield ground-state masses that deviate by less than $0.5\%$ from the highly precise Particle Data Group (PDG) experimental averages (5.279, 5.366, and 6.274 GeV, respectively). The calculated mass spectra exhibit high parametric stability, with standard deviations of $0.028$~GeV and $0.031$~GeV for the Numerov and variational methods, respectively. This accuracy in mass computation indicates that the Gaussian trial wavefunction provides an excellent mathematical fit for the spatial distribution. Fundamentally, this arises because a harmonic-oscillator term in the potential ($ar^2$) yields a Gaussian function as the exact ground-state wavefunction. For the $B$ meson, both approaches systematically overestimate the leptonic decay constant, yielding $0.365$~GeV (Numerov) and $0.314$~GeV (variational), which exceed experimental values by approximately $84\%$ and $58\%$, respectively. On the other hand, for the $B_s$ meson, the calculated decay constants demonstrate improved alignment with experimental measurements, where the upper bound of the Numerov average ($0.176\pm0.009$~GeV) deviates by roughly  $15\%$ from the experimental data. On the other hand, the lower bound of the variational average ($0.254GeV$) deviates by around $10\%$ from the lower bound experimental average. For the double-heavy $B_c$ meson, the calculated upper bounds for the decay constant show strong agreement with existing theoretical literature. This proves that our results for the decay constants demonstrate improved accuracy for systems in which both the quark and antiquark is heavier. For the $B(\bar{b}d)$ meson, the $d$ quark ($0.336\text{ GeV}$) is exceedingly light compared to the bottom quark ($4.95\text{ GeV}$). In the case of the $B_s$ meson, however, the $s$ quark is comparatively heavier than the $d$ quark in the $B$ meson. Consequently, the presence of a heavier light-sector quark makes this nonrelativistic framework more applicable. Nevertheless, for extremely heavy-light configurations, relativistic treatments must be employed. Also, because the neutral meson mass difference ($\Delta m_B$) is quadratically dependent on the decay constant ($f_p$), it serves as an exceptional tool in non-relativistic bound-state frameworks. The systematic overestimation in $B_d$ and the dual bracketing behaviours in the $B_s$ system serve as a clear physical indicator of the spatial limits of the Killingbeck potential.

The primary volatility in this theoretical framework lies in the evaluation of the short-distance wavefunction. The overestimation of the probability density at the origin ($|\Psi(0)|^2$) stems from a highly rigid, localised potential at $r=0$. We conclude that resolving this discrepancy requires the incorporation of short-range relativistic corrections, the Darwin term\cite{Buisseret_2007} and the mock-meson mass\cite{vwg1-nz37} formulation. The Darwin term accounts for the rapid quantum fluctuations of the light quark in the $B$ and $B_s$ systems, ensuring the quark experiences a smeared average of the central potential rather than a singular interaction at a mathematical point $r=0$. Furthermore, the mock mass approach redefines the system as a dynamic ``mock meson'' state, mitigating the physical limitations of treating the heavy-light system as a purely static, non-relativistic bound state.
Our calculated slope for the $B$ meson ($\rho^{2} \approx 0.338$--$0.339$) is lower than our previous work and below the theoretical bounds reported by Le Yaouanc \textit{et al.} and Das \textit{et al.} It comfortably satisfies the model-independent Bjorken lower bound, $\rho^{2} \geq \tfrac{1}{4}$\cite{oliver2014isgur}, confirming the basic consistency of our wavefunction-overlap calculation, but it falls short of the stronger Uraltsev bound\cite{uraltsev2001new}, $\rho^{2} \geq \tfrac{3}{4}$, which is expected to hold in the strict heavy-quark limit. This shortfall is consistent with the fact that our treatment is purely non-relativistic: our $B$-meson value lies intermediate between Hazarika \textit{et al.}'s non-relativistic $\overline{\mathrm{MS}}$- and $V$-scheme results ($0.416^{\mathrm{a1}}$, $0.431^{\mathrm{a2}}$) and their relativistically corrected counterparts ($0.258^{\mathrm{b1}}$, $0.244^{\mathrm{b2}}$), suggesting that the inclusion of relativistic effects would be required to approach both the Uraltsev bound and full agreement with comparable potential-model calculations. Our curvature ($C \approx 0.032$--$0.033$), while lower than our previous work, Le Yaouanc \textit{et al.}, and Das \textit{et al.}, is in fact $5$--$8$ times larger than all four of Hazarika \textit{et al.}'s curvature values ($0.0040$--$0.0071$), reinforcing that relativistic corrections---which act to suppress the curvature more strongly than the slope in Hazarika's scheme-by-scheme comparison---are the most probable source of this discrepancy, rather than the confining potential's functional form.

For the $B_{s}$ meson, our calculated slope ($\rho^{2} \approx 0.741$) also lies just below the Uraltsev bound, $\rho^{2} \geq \tfrac{3}{4}$, and shows close agreement with Hazarika \textit{et al.}'s non-relativistic $\overline{\mathrm{MS}}$-scheme result ($0.733^{\mathrm{a1}}$), while remaining lower than all other literature values reported in Table~\ref{tab:slope_comparison_all}. In the $B_{c}$ sector, by contrast, both constituent quarks are heavy, and the system undergoes a pronounced spatial collapse into an ultra-compact, tightly bound configuration. According to the uncertainty principle, this strong localisation in position space corresponds to a broad momentum-space distribution, rendering the bound-state wavefunction highly sensitive to even small changes in the heavy-quark velocity during a semileptonic transition. The resulting mismatch between the initial- and final-state momentum-space wavefunctions leads to a rapid reduction in their quantum overlap, manifesting as a steeply decreasing Isgur--Wise function. This behavior is consistent with our significantly larger values of the slope, $\rho^{2} \approx 3.70$, and the curvature, $C \approx 3.8$--$4.0$, for the $B_{c}$ meson. The same physical mechanism operates at the opposite extreme for the $B$ meson: its comparatively small values of $\rho^{2}$ and $C$ reflect an extended, loosely bound valence-quark distribution with a correspondingly narrow momentum-space profile and reduced sensitivity to recoil, whereas the compact and rigid binding of the $B_{c}$ meson gives rise to the opposite behaviour.
However, our non-relativistic framework, implemented using both the Matrix Numerov and variational methods with the Killingbeck potential, yields closely consistent meson mass spectra. Nevertheless, the Killingbeck potential is expected to be more effective for heavy--light systems, such as the $B$ and $B_s$ mesons, when relativistic corrections are incorporated, as discussed earlier. Furthermore, the Killingbeck potential may provide a suitable description of quark--antiquark interactions in environments where color confinement is modified, such as the quark--gluon plasma (QGP). Under appropriate approximations, the screened Coulomb potential reduces to the Killingbeck form, with an additional temperature-dependent quadratic term, $a(T)r^{2}$, where the coefficient $a(T)$ depends on the medium temperature. Consequently, the present study motivates future investigations of the evolution of meson bound states in high-temperature environments such as the QGP, including the transition from bound to unbound states and the determination of the corresponding dissociation (critical) temperatures under such extreme conditions.

\section*{Appendix A: Physical Justification and Derivation of the Harmonic Term}
\label{app:harmonic_derivation}

Our motivation stems from the fact that heavy quarkonium states interact with a complex non-perturbative vacuum populated by soft gluonic background fluctuations and vacuum polarization effects. Thus, thermal medium screening within a spectroscopic framework is physically grounded for this case.  Mathematically, these effective vacuum dynamics can be mapped onto a finite-temperature description via the Karsch-Mehr-Satz (KMS) screened potential \cite{Karsch1988}. Thermalising the extended Cornell potential provides a field-theoretic basis for the harmonic $a'r^2$ term, replacing abstract fitting coefficients with fundamental QCD observables ($\alpha_s(T)$, $b(T)$, and $m_D(T)$) and bridging zero-temperature spectroscopy with high-density environments.

The finite-temperature KMS potential incorporating Debye screening for both vector Coulomb exchange and scalar string confinement is defined as:
\begin{equation}
V(r, T) = -\frac{\alpha_s(T)}{r} e^{-m_D(T) r} + b(T) \, r \, e^{-m_D(T) r}.
\label{eq:kms_potential}
\end{equation}

Within the bound-state spatial regime where $m_D r \ll 1$, expanding the screening exponential factor $e^{-m_{D} r}$ using a Taylor series gives:
\begin{equation}
e^{-m_D r} = 1 - m_D r + \frac{m_D^2 r^2}{2} - \frac{m_D^3 r^3}{6} + \mathcal{O}(r^4).
\label{eq:taylor_expansion}
\end{equation}

Applying Eq.~\eqref{eq:taylor_expansion} separately to the screened Coulombic and confining terms yields the following:
\begin{align}
V_{\text{Coul}}(r, T) &= -\frac{\alpha_s}{r} \left( 1 - m_D r + \frac{m_D^2 r^2}{2} - \frac{m_D^3 r^3}{6} + \mathcal{O}(r^4) \right) \nonumber \\
&= -\frac{\alpha_s}{r} + \alpha_s m_D - \frac{\alpha_s m_D^2}{2} r + \frac{\alpha_s m_D^3}{6} r^2 + \mathcal{O}(r^3), \label{eq:v_coul_expanded} \\
V_{\text{conf}}(r, T) &= b r \left( 1 - m_D r + \frac{m_D^2 r^2}{2} + \mathcal{O}(r^3) \right) \nonumber \\
&= b r - b m_D r^2 + \frac{b m_D^2}{2} r^3 + \mathcal{O}(r^4). \label{eq:v_conf_expanded}
\end{align}

Summing Eqs.~\eqref{eq:v_coul_expanded} and \eqref{eq:v_conf_expanded} and truncating the series in order $\mathcal{O}(r^2)$ results in:
\begin{equation}
\begin{split}
V(r, T) ={} & -\frac{\alpha_s}{r} + \alpha_s m_D + \left( b - \frac{\alpha_s m_D^2}{2} \right) r \\
& + \left( \frac{\alpha_s m_D^3}{6} - b m_D \right) r^2 + \mathcal{O}(r^3).
\end{split}
\label{eq:v_total}
\end{equation}

Comparing Eq.~\eqref{eq:v_total} term by term with the phenomenological extended Killingbeck potential,
\begin{equation}
V_{\text{Kill}}(r) = a' r^2 + b' r + \frac{c'}{r} + d',
\label{eq:killingbeck_standard}
\end{equation}
establishes an exact functional mapping for the parameters:
\begin{align}
a' &= \frac{\alpha_s(T) \, m_D^3(T)}{6} - b(T) \, m_D(T), \label{eq:param_a} \\
b' &= b(T) - \frac{\alpha_s(T) \, m_D^2(T)}{2}, \label{eq:param_b} \\
c' &= -\alpha_s(T), \label{eq:param_c} \\
d' &= \alpha_s(T) \, m_D(T). \label{eq:param_d}
\end{align}

We see that rather than acting as an ad-hoc fit parameter, $a'$ reflects the interplay between short-range screening corrections ($\frac{\alpha_s m_D^3}{6} > 0$) and thermal attenuation of confinement ($-b m_D < 0$), as shown in Eq.~\eqref{eq:param_a}. The $a' r^2$ harmonic term thereby establishes a rigorous theoretical link between in-medium gluonic dynamics and heavy-quark confinement properties.

\bibliography{apssamp}

\end{document}